\documentclass{mn2e}
\usepackage{psfig}
\usepackage{longtable}

\def\ltsima{$\; \buildrel < \over \sim \;$}
\def\lsim{\lower.5ex\hbox{\ltsima}}
\def\gtsima{$\; \buildrel > \over \sim \;$}
\def\gsim{\lower.5ex\hbox{\gtsima}}

\def\ltsima{$\; \buildrel < \over \sim \;$}
\def\simlt{\lower.5ex\hbox{\ltsima}} 
\def\gtsima{$\; \buildrel > \over \sim \;$}
\def\simgt{\lower.5ex\hbox{\gtsima}} 
\begin{document}

\title[Cluster-short GRB correlation]{On the correlation of Short
Gamma--Ray Bursts and Clusters of galaxies}

\author[Ghirlanda et al.]
{G. Ghirlanda$^1$, M. Magliocchetti$^2$, G. Ghisellini$^1$, L. Guzzo$^1$ \\
$^1$ INAF, Osservatorio Astronomico di Brera, via E. Bianchi 46,
I-23807
Merate (LC), Italy {\tt e-mail: ghirla@merate.mi.astro.it}\\
$^2$ SISSA/ISAS, Via Beirut 2-4, I-34014 Trieste, Italy\\ 
}

\maketitle

\begin{abstract}
We cross correlate Gamma--Ray Bursts and X--Ray selected clusters of
galaxies at $z\leq0.45$. We find a positive $2\sigma$ signal for the
angular cross--correlation function $w_{bc}(\theta)$ on scales
$\theta\leq3^\circ$ between short GRBs and clusters. Conversely, no
correlation is found between clusters and the population of long GRBs.
The comparison with the cluster autocorrelation function shows that
short GRBs do not trace the cluster distribution as not all short GRBs
are found in clusters.  A higher signal in $w_{bc}(\theta)$ is found
if we only consider the cluster population up to $z$=0.1.  By
comparing the short burst autocorrelation function with model
predictions we then constrain short bursts to mostly originate within
$\sim 270$ Mpc (i.e. $z\leq 0.06$). Our analysis also reveals that
short GRBs are better correlated with ``normal'' galaxies. The double
compact object merger model for short GRBs would associate them
preferentially to early--type galaxies but the present statistics do
not allow us to exclude that at least a fraction of these events might
also take place in late--type galaxies, in agreement with recent
evidences.
\end{abstract}

\begin{keywords}
gamma-ray: bursts --- methods: data analysis 
\end{keywords}

\section{Introduction}

The population of Gamma Ray Bursts (GRB) presents a bimodal duration
distribution (Koveliotu et al. 1999): short GRBs (S--GRBs - lasting
less than 2 s) and and long GRB (L--GRBs - lasting more than 2
s). Further support to this bimodality comes from their different
spectral properties with short GRBs being harder than longer events
(Tavani 1998; Ghirlanda, Ghisellini \& Celotti 2004).

One of the most accredited models associates short GRBs to the
coalescence and final merger of two compact objects in a binary system
(Narayan, Paczynski \& Piran 1992; Ruffert 1997). Binary systems
evolve on different timescales (Voss \& Tauris 2003) and a testable
consequence of this model is that short GRBs should be found
preferentially in evolved (early) type galaxies.  Recently, {\it
Swift} (Gehrels et al. 2004) and {\it Hete--II} (Lamb et al. 2004)
detected for the first time the X--ray and optical afterglow emission
of three short GRBs and in two cases a redshift was measured:
GRB~050724 (Covino et al. 2005) was found to be associated with an
elliptical galaxy at $z=0.257$ (Berger et al. 2005) whereas GRB~050709
(Butler et al. 2005) was found to be associated with a blue dwarf
galaxy at redshift $z=0.16$ (Covino et al. 2005, Hjorth et al. 2005)

A third interesting case is represented by GRB~050509B (Hurkett et
al. 2005), whose afterglow was detected only in X--rays, was found to
be spatially coincident with a giant elliptical galaxy at $z=0.2248$
(Bloom et al. 2005) belonging to the cluster of galaxies ZwCl
1234.0+02916 at $z=0.2214$ (Pedersen et al. 2005). Another short
GRB~050813 was found to be spatially coincident with a cluster at
$z=0.7$ (Gladders et al. 2005). Gal--Yam et al. (2005) recently
reported a third significant positional coincidence of the short
GRB~790613 with the rich Abell cluster 1892 at $z=0.09$.  The
possibility of a correlation of GRBs with clusters of galaxies has
been debated in the past (Kolatt \& Piran 1996, Struble \& Rood 1997)
by studying the direct and statistical association of BATSE bursts
with optically selected Abell clusters, and opposite conclusions were
reached (Hurley et al. 1999; Gorosabel \& Castro--Tirado 1997). In
these studies, however, the population of bursts was not separated in
short and long events.  Magliocchetti, Ghirlanda \& Celotti (2003,
hereafter MGC03) found evidence for anisotropy in the distribution of
short GRBs and suggested that they might originate in galaxies
distributed up to $z\sim 0.5$. Tanvir et al. 2005 found that at least
a fraction between 5\% and 25\% of BATSE short GRBs might be in the
very local universe (i.e. $z<0.025$) preferentially in early type
galaxies.  It has also been proposed that a fraction (e.g. Hurley et
al. 2005) of short GRBs might be the cosmological counterparts of the
giant flares of Soft Gamma Ray Repeaters (SGR).

In this letter we study the cross correlation between short GRBs
detected by BATSE and X--ray selected clusters of galaxies covering
the entire sky ($\pm$20$^\circ$ above and below the Galactic plane). We
find that a positive correlation indeed exists between the two
population of objects on small angular scales. We investigate if short
GRBs trace the cluster distribution and also test if they
preferentially correlate with early type galaxies.  In the following
we assume a ``standard'' cosmological model with $\Omega_{M}=0.3$ and
$\Omega_{\Lambda}=h=0.7$.
\begin{figure}
\vskip -0.3 true cm
\begin{center}
\psfig{file=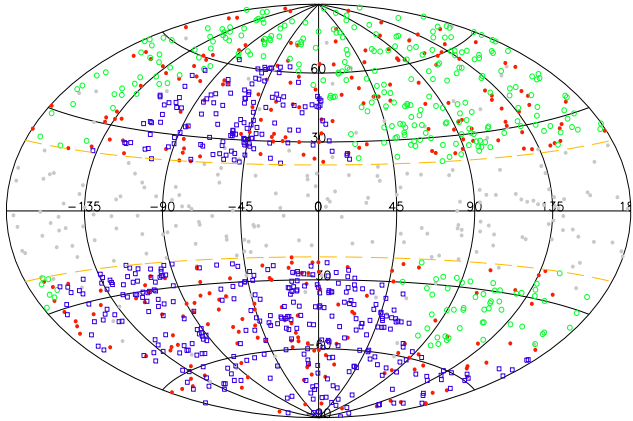,width=8.5cm,height=6cm}
\caption{Sky distribution in Galactic coordinates of the sample of 497
BATSE short GRBs (filled circles) and of 763 Clusters (449 REFLEX
clusters - blue open circles - and 314 NORAS clusters - green open
circles). The red filled circles represent the 283 short duration GRBs
with position accuracy $<10^{\circ}$ and $|b|>20^{\circ}$.\label{skydist} }
\end{center}\vskip -0.8 true cm
\end{figure}
\section{GRBs and Clusters samples}

BATSE detected more than 2000 GRBs during its nine years activity
(Paciesas et al. 1999).  From the on-line sample of
BATSE--GRBs\footnote{http://cossc.gsfc.nasa.gov/docs/cgro/batse/} we
extracted 497 short events with duration $\leq2$ s and 1540 long
GRBs with duration $>2$ s. 
\begin{figure}
\vskip -0.3 true cm
\begin{center}
\psfig{file=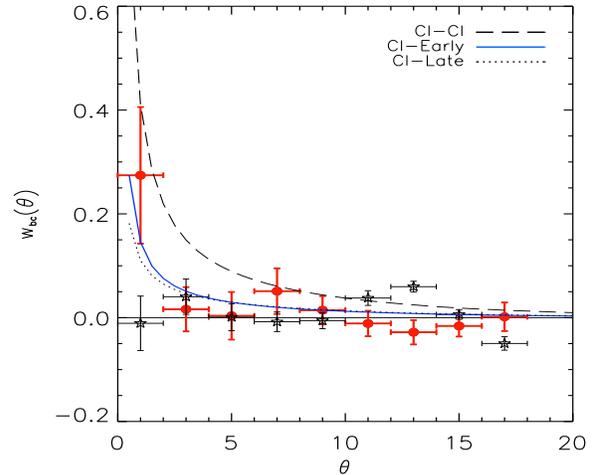,width=8.5cm,height=7cm}
\caption{Angular cross-correlation (red filled circles) between short GRBs
and X---ray selected clusters of galaxies . Also shown is (open stars)
the signal obtained with BATSE long GRBs . The long--dashed line
represents the autocorrelation function of clusters, while the solid
and dotted lines represent the cross correlation of clusters with
early type and late type galaxies (see Sec. 3 for details).
\label{fig2}
}
\end{center}\vskip -0.8 true cm
\end{figure}
In order to correlate the GRB sample with a sample of clusters of
galaxies with a sky coverage as wide as possible, we considered the
REFLEX (Boehringer et al. 2004) and the NORAS (Boehringer et al. 2000)
surveys.  The {\it ROSAT} ESO Flux Limited X--ray (REFLEX) galaxy
cluster survey contains 449 clusters with a maximum redshift $z\sim
0.451$ and it is flux limited to 3$\times 10^{-12}$ erg cm$^2$
s$^{-1}$ in the [0.1-2.4] keV band. It covers the Galactic latitudes
$|b|\geq20^\circ$ and declination $\delta\leq2.5^\circ$ and it
excludes $\sim$ 324 deg$^2$ of sky around the LMC and SMC. The total
area covered is 4.24 sr (i.e. $\sim$ 34\% of the entire sky) and the
catalog completeness is $\geq 90$\% (Boehringer et al. 2004).  The
Northern $ROSAT$ All-Sky galaxy survey (NORAS) contains 484 clusters
(including the supplements to the original survey catalogue -
Bohringer et al. 2000) with measured redshift up to $z\sim 0.459$. It
covers the Galactic latitudes $|b|\geq20^\circ$ and declination
$\delta\geq0^\circ$.  We are aware that the NORAS sample is only 82\%
complete with respect to the REFLEX survey and we tested our results
for this difference (Sec. 3). In combining the two surveys, given the
higher level of completeness of REFLEX, we cut NORAS at $\delta\geq
2.5^\circ$, to avoid superpositions, and to the same flux limit of
REFLEX. We end up with 314 (out of 484) NORAS clusters combined with
the 449 REFLEX clusters (open green and blue circles in
Fig.~\ref{skydist}, respectively).

Since we do not have redshift measurements for the population of short
GRBs we have to rely on projected quantities. Note that for the cross
correlation analysis we have considered only the 283 (out of 497)
short GRBs (red filled circles in Fig.~\ref{skydist}) distributed at
Galactic latitudes $|b|\geq20^\circ$ and with positional accuracy
$\Delta<10^\circ$. The latter selection limits the large uncertainties
associated to the position of BATSE bursts (see MGC03 and Sec.~3). The same
selection applied to the sample of 1540 long bursts gives 973 events.
In Fig.~\ref{skydist} we report the final samples of short GRBs
(filled circles) and the two samples of clusters (open symbols).

\section{GRBs--Clusters cross correlation}

The angular cross correlation $w_{bc}(\theta)$ represents the
fractional increase, relative to a random distribution, of the
probability per unit solid angle of finding objects separated by an
angle $\theta$. We computed $w_{bc}(\theta)$ for the 283 S--GRB (B)
and the 762 clusters (C) by counting the burst--cluster pairs (BC) at
different angular separations $\theta$. This is compared with the
pairs found at the same angular scale $\theta$ between the burst
sample and a random sample (BR) of $\sim 10^{5}$ objects distributed
uniformly in the same sky portion covered by the clusters samples.

We use the Hamilton  estimator for the angular cross correlation
function between short GRBs and clusters (Hamilton 1993):
\begin{equation}
w_{bc}(\theta)=\frac{N_{BR}^2}{N_{BC}N_{RR}}\frac{BC(\theta)RR(\theta)}{BR^2(\theta)} -1
\end{equation}
where $N_{BR}$ and $N_{BC}$ represent the total number of pairs
between the short burst sample and the random and cluster samples,
respectively, and $N_{RR}$ represents the number of pairs in the
random sample. The errors on $w_{bc}(\theta)$ were derived with the
bootstrap method (see e.g. Ling, Frenk \& Barrow 1986): a set of 100
bootstrap catalogs, each the same size as the data catalog, were
randomly extracted from the GRB catalog. The cross--correlation
(Eq.~1) was computed for each bootstrap catalog and, for each
$\theta$, a set of normally distributed estimates of the correlation
function is obtained. The variance around the mean represents the
1$\sigma$ uncertainty on $w_{bc}(\theta)$.  In all our calculation we
accounted for the BATSE sky exposure map (Chen \& Hakkila 1997) and
for the sky exposure of the REFLEX and NORAS samples, which however
are quite uniform. As already shown in Fig.~1 of MGC03, the
typical positional error associated with short BATSE bursts is of few
degrees and the distribution of positional uncertainties of short
bursts is nearly flat up to $\sim 5^\circ$ while that of long GRBs
peaks at 1-2$^\circ$.  The large uncertainty in the positions of short
GRBs might affect the correlation results. Following the same
procedure adopted in MGC03 we tested how the cross-correlation signal
changes by considering subsamples of short GRBs with better positional
determination (i.e. $\leq5^\circ$) and found results fully consistent
with those presented in Fig.~\ref{fig2}.

We find a positive $2\sigma$ correlation signal (Fig.~\ref{fig2} - red
filled circles) on small angular scales (i.e. $\theta<3^\circ$)
between short GRBs and clusters while no correlation is found between
long GRBs and clusters (star symbols in Fig.~\ref{fig2}).

Given the incompletness of the NORAS cluster sample (Sec.~2), we
checked our results by computing the cross correlation between S-GRBs
and the REFLEX and NORAS samples separately. Although the signal has
larger uncertainties due to the smaller number of objects in the
individual samples, we still find a positive correlation signal on
small angular scales, entirely compatible with the one obtained by
considering the two samples together.

\section{Model comparison}

To provide an insight on the findings of the previous section we have
to compare our data with results available in the literature.  We use
the generalization of the Limber equation developed by Peebles (1974)
and Lilje \& Efstathiou (1988), which relates the angular two-point
cross-correlation function $w_{c,i}(\theta)$ to the spatial one,
$\xi_{c,i}(r)$:
\begin{eqnarray}
w_{c,i}(\theta)=2\cdot\frac{\int_0^\infty N_c(x)\Phi_i(x) x^2 dx
\int_0^\infty \xi_{c,i}(r) du}
{\int_0^\infty N_c(x)dx\int_0^\infty \Phi_i(x) x^2 dx},
\label{eq:limber}
\end{eqnarray}
where $x$ is the comoving coordinate, $u$ and $r$ are related by the
expression (which only holds in the small angle approximation,
$\theta<< 1$~rad) $r^2\simeq u^2+\theta^2x^2$, $N_c(x)$ is the number
of clusters in the considered sample with distance between $x$ and
$x+dx$ and the selection function for a particular class of
extragalactic sources $\Phi_i(x)$ is connected to their redshift
distribution by
\begin{eqnarray}
\int \Phi_i(x) x^2 dx=\frac{1}{\omega_s}\int N_i(z) dz, 
\end{eqnarray}
where $\omega_s$ is the  area covered by the survey. Note that 
the above equations have been obtained for a flat universe.
 
For our analysis, we chose to consider three different cases for the
spatial cross-correlation function: (i) $\xi_{c,i}\equiv \xi_{c,c}$
(i.e. cross-correlation function coinciding with the
auto-correlation function of clusters); (ii)
$\xi_{c,i}\equiv\xi_{c,e}$ (cross-correlation between clusters and
early-type galaxies); (iii) $\xi_{c,i}\equiv\xi_{c,l}$
(cross-correlation between clusters and late-type galaxies).

As for the form of the cross-correlation function to be plugged into
equation (\ref{eq:limber}), in (i) we have used $\xi_{c,c}(r)=\left(
\frac{r}{r_0}\right)^{-\gamma}$, with $r_0=18.8/h$~Mpc and
$\gamma=1.83$, as derived from the analysis of the clustering
properties of REFLEX clusters (Collins et al. 2000), while for the
other two cases (ii,iii) we have written
$\xi_{c,i}(r)=\xi_{c,c}\;b_c/b_i$, with $b$ the bias factor. This
latter expression can be easily derived as, theoretically, the
cross-correlation function of two populations of extragalactic sources
can be expressed as $\xi_{j,i}=b_j b_i\xi_{\rm DM}$, with $\xi_{\rm
DM}$ autocorrelation function of the underlying dark matter (see
e.g. Mo, Peacock \& Xia, 1993).  Under the assumption of
scale-independence, $b_i=\sigma_{8,i}/\sigma_{8}$, where $\sigma_8$ is
the r.m.s. of matter fluctuations on a scale $8/h$~Mpc, and
$\sigma_{8,i}$ can be obtained from measurements of the two-point
autocorrelation function for a chosen class of sources as
$\sigma_{8,i}=\left[c_{\gamma_i}\;\left(r_{0,i}/8\right)^{\gamma_i}\right]
^{0.5}$, with
$c_{\gamma_i}=\frac{72}{(3-\gamma_i)(4-\gamma_i)(6-\gamma_i)
\;2^{\gamma_i}}$ (Peebles, 1980). Values for the correlation length
$r_{0,i}$ and the slope $\gamma_i$ for the populations of early-type
and late-type galaxies have been taken from studies of the clustering
properties of 2dF galaxies (Madgwick et al., 2003) and, for
$\sigma_8=0.9$ as the latest results from CMB data seem to indicate
(Spergel et al. 2003), we obtain $b_c=4.05$, $b_e=1.54$, $b_l=0.97$
respectively for the bias factor of clusters, early-type and late-type
galaxies.

$N_c(x)$ in equation (\ref{eq:limber}) has been derived from the
redshift distribution of REFLEX clusters. Note that in this case we
have also made the choice $N_c(z)\equiv N_i(z)$, i.e. we have assumed
galaxies of both types to follow the redshift distribution of REFLEX
clusters.  This choice was forced by the lack of a statistically
significant spectroscopic sample of galaxies of different type which
probes their redshift evolution from the local universe up to $z\simeq
0.45$, maximum redshift of the clusters analysed in this work.

\section{Results}

Direct comparison with the data shows that the distribution of
short-GRBs (red filled circles in Fig.~\ref{fig2}) does not trace that of
clusters (long dashed line in Fig.\ref{fig2}), i.e. there is not a strong
correspondence between clusters and short-GRBs, as not all short-GRBs
are found to reside in clusters. 
If one instead compares the results with the cluster-galaxies
cross-correlation function (solid and dotted lines in Fig.~\ref{fig2}
respectively for early-type and late-type galaxies), a much better
agreement is obtained.

This implies that short-GRBs exhibit similar clustering properties of
``normal'' galaxies, i.e. that they are present in these classes of
sources. Unfortunately, the size of the error-bars does not allow to
discern between early-type and late-type galaxies, preventing to
associate short-GRBs with a particular class of galaxies.
\begin{figure}
\vskip -0.3 true cm
\begin{center}
\psfig{file=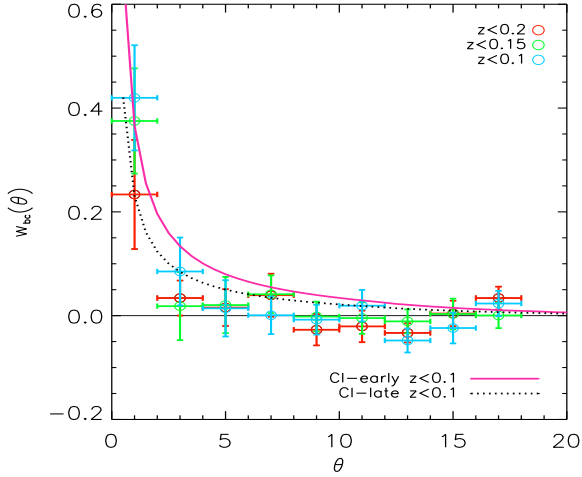,width=8.5cm,height=7.cm}
\caption{Cross correlation between short GRBs and redshift limited
samples of clusters of galaxies ($z< 0.1,0.15,0.2$). The solid
(dotted) line represents the cross correlation between clusters and
early type (late type) galaxies at redshifts $z\leq0.1$.
\label{fig3}
}
\end{center}\vskip -0.5 true cm
\end{figure}
\begin{figure}
\vskip -0.3 true cm
\begin{center}
\psfig{file=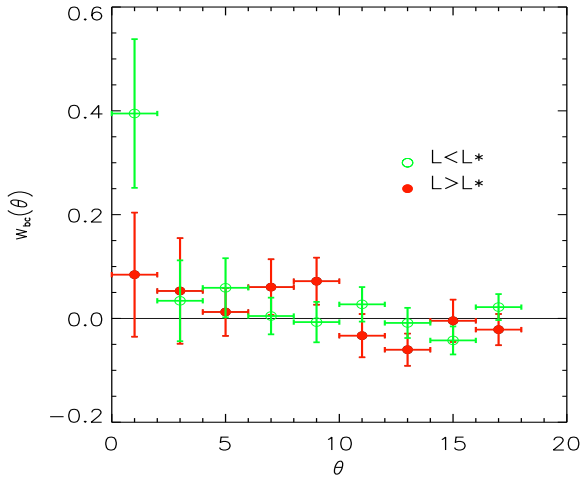,width=8.5cm,height=7.cm}
\caption{Cross correlation function of the sample of short GRBs and
the clusters of galaxies sample separated in two luminosity
sub-samples of roughly equal total number of objects. The open symbols
represent the galaxy clusters with X-ray luminosity
$<1.4\times10^{44}$ erg/s ($<1.9\times10^{44}$ erg/s) for the
REFLEX (NORAS) sample.
\label{fig6}
}
\end{center}\vskip -0.8 true cm
\end{figure}

The local population of short bursts might be ``contaminated'' by a
more distant (i.e. $z>0.5$) and isotropic population of objects which
reduces the correlation signal with clusters at $z<0.5$.  Under this
hypothesis we should expect the cross correlation signal to decrease
when selecting sub--samples of clusters at lower redshifts.

We tested for this possibility by cutting the REFLEX+NORAS cluster
sample at different redshifts. At variance with what previously
expected we find (Fig.~\ref{fig3}) a {\it stronger} correlation
signal between S--GRBs and the 420 clusters with $z\leq0.1$ than what
found with the 693 clusters distributed out to $z=0.2$ or out to
$z\sim0.4$.  We also tested for any dependence of the
cross-correlation signal on the cluster luminosity. In order to do
this, we have divided the REFLEX and NORAS cluster samples in two
sub--samples by considering the median values of their X--ray
luminosity distribution: $L_{X}\sim 1.94 \times 10^{44}$ erg s$^{-1}$
and $L_{X}\sim 1.37 \times 10^{44}$ erg s$^{-1}$ for the NORAS and
REFLEX surveys, respectively. This selection corresponds to have a
roughly equal number of clusters ($\sim$424) in the two luminosity
samples. As it is shown in Fig.~\ref{fig6}, we find a higher signal
with sub-luminous clsuters with respect to the signal found with more
luminous clusters\footnote{Given the higher signal found with clusters
at $z\leq0.1$ we also tried to apply both the luminosity and redshift
cuts. With these combined cuts our results are strongly affected by
the paucity of clusters in the more luminous bin (e.g. only 78 with
$L>L_{*}$ and $z<0.1$) which does not allow us to draw any
conclusion.}.

Since the cluster surveys considered in our work are both
flux-limited, one has that in general more luminous sources will
mainly be more local than sub-luminous ones. What the data then shows
is that short-GRBs might preferentially inhabit low-redshift ($z\simlt
0.1$) systems.  In fact, this evidence is furtherly strengthened by
the comparison of our results with the angular cross-correlation
function of clusters and galaxies, obtained as in equation
(\ref{eq:limber}) by applying a redshift cut $z_{\rm max}=0.1$.  One
can see (Fig.~\ref{fig3}) that there is a very good match between data
and ``models''. Note that in this case the redshift distribution
adopted for early-type and late-type galaxies has been taken in a
self-consistent way from the 2dF Galaxy Redshift Survey (Madgwick et
al. 2002), complete at least up to $z\simeq 0.15$.
\begin{figure}
\vskip -0.3 true cm
\begin{center}
\psfig{file=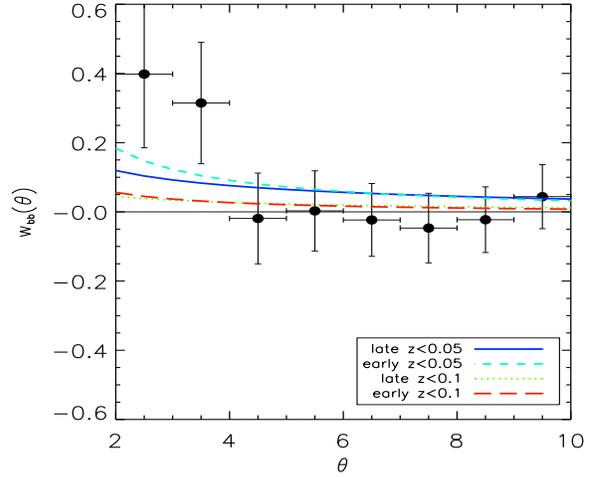,width=8.5cm,height=7.cm}
\caption{Two point angular auto-correlation function for the short GRB
sample with positional error $\Delta<10^\circ$. The lines represent
different autocorrelation functions for early and late type galaxies
with limiting redshift of 0.1 and 0.05.
\label{fig5}
}
\end{center}\vskip -0.8 true cm
\end{figure}

Further evidence on the proximity of short-GRBs can be obtained by
comparing the projected auto-correlation function for this class of
sources as obtained by MGC03, with the autocorrelation function of
both late-type and early-type galaxies evaluated in a fashion similar
to equation (\ref{eq:limber}) up to some maximum redshift $z_{\rm
max}$. This is done in Fig.~\ref{fig5}, where the dotted (long dashed)
line represents the autocorrelation function of late (early) type
galaxies at $z\leq0.1$. The solid (dashed) line is instead that for
$z\leq0.05$. This comparison with the data also seems to hint to
short-GRBs to be more local than $z\le 0.1$, possibly being
distributed only up to redshifts $z=0.05$. Again we note that, with
the available data, it is not possible to discern if short-GRBs are
mainly hosted in early-type or late-type galaxies. Better data
(probably samples of size about three-to-five times the one considered
in this work) are needed to draw more firm conclusions.

At face value, these results point to short GRBs to be local and
preferentially associated to less dense environments (i.e., by
extrapolation, normal galaxies) rather than to those identified by the
clusters themselves. This last conclusion is reached as less
X-ray-luminous clusters are generally associated to less massive
systems (see e.g. Borgani et al., 2002).

A consequence of these results is that, given the average fluence
(integrated $>25$ keV) of 4.3$\times 10^{-7}$ erg cm$^{-2}$ s$^{-1}$
of our sample of short GRBs, we estimate a typical isotropic
equivalent energy of $\sim 2.4\times10^{48}$ erg, by assuming
$z=0.05$. Another consequence of having short GRBs at lower redshift
than previously thought is the increase of their rate as recently
pointed out by Guetta \& Piran 2005, Nakar et al. 2005 and Gal--Yam et
al. 2005.

\section{Conclusions}

By cross correlating BATSE bursts and (REFLEX + NORAS) X--ray selected
samples of clusters (distributed out to $z\simeq 0.45$) we find that
short duration ($T_{90}<2$ s) GRBs are correlated with clusters
while we do not find any correlation with the population of long GRBs.
By comparing the short GRBs--cluster correlation signal with the
cluster--cluster autocorrelation signal we can exclude that short
bursts trace the cluster distribution. Instead, through the comparison
with the cluster--galaxy correlation functions we conclude that short
GRBs are associated to ``normal'' galaxies.  

We explored the hypothesis that short GRBs are local events. In fact,
we find a higher cross--correlation signal with low redshift clusters
(i.e. up to $z\leq0.06$) or with sub-luminous clusters. What furtherly
supports our conclusion is (i) the similarity of the short
GRBs-clusters and local galaxies-clusters cross correlation and (ii)
the short GRBs auto-correlation, which is similar to the
auto--correlation function of local ($z\leq$0.1) galaxies.  The
present statistics do not allow to exclusively associate short GRBs
with early--type galaxies as expected if they are produced - in the
double compact merger scenario - by an old stellar population.

These results represent a challenge because on one side they seem to
contradict the, still very few, redshift measurements of short GRBs
which place them at $z=0.16$ (GRB~050709 - Covino et al. 2005, Hjorth
et al. 2005) and $z=0.25$ (GRB~050724 - Berger et al.  2005) and on
the other side they predict a typical energy of $\sim10^{48}$ erg for
short GRBs.  The apparently different (lower) redshift of BATSE
short bursts (as found with our analysis, but see also Tanvir et
al. 2005) with respect to the few Swift measured redshifts might be
due to one (or to a combination) of several effects: (i) an unknown
bias of Swift to detect systematically larger redshifts (as also shown
for the population of long GRBs) with respect to BATSE; (ii) a
possible contamination of the BATSE short burst population by
extragalactic SGR (as suggested by Hurley et al. 2005), although
direct positional searches (Nakar et al. 2005) and spectral analysis
(Lazzati et al. 2005) hardly support this scenario; (iii) a complex
luminosity function (e.g. Guetta \& Piran 2005).  For these reasons
the redshift distribution of short bursts still represents an open
issue which needs more direct redshift determinations as well as a
better understanding of the possible selection effects.

\section*{Acknowledgments}
We thank A. Celotti, F. Tavecchio, S. Molendi and S. De Grandi for
useful discussions. We thank the referee for her/his useful
comments. We acknowledge MIUR for funding (Cofin grant
2003020775\_002).


\begin{thebibliography}{}
\bibitem{} Berger E. et al., 2005, Nature subm.,  astro-ph/0508115
\bibitem{} Bloom J. S. et al, 2005, ApJ in press, astro-ph/0505480
\bibitem{} Bohringer H. et al., 2000, ApJSS, 129, 435
\bibitem{} Bohringer H. et al., 2004, A\&A, 425, 367
\bibitem{} Borgani S., Governato F., Wadsley J., Menci N., Tozzi P., 
Quinn T., Stadel J., Lake G., 2002, MNRAS, 336, 409. 
\bibitem{} Butler et al., 2005, GCN 3570
\bibitem{} Cline B. D., Matthey C., Otwinowski S., 1999, ApJ, 527, 827
\bibitem{} Covino S. et al., 2005, GCN, 3665
\bibitem{} Djorgovski S. G., et al., 1997, IAU Circ. 6660 
\bibitem[\protect\citename{collins }2000]{co}
Collins C.A., 2000MNRAS.319..939C
\bibitem{} Kawai et al., 2002, Gamma--Ray Bursts and Afterglow
Astronomy, eds. G. R. Ricker and R. Vanderspek (New York: AIP), 224
\bibitem {} Kolatt T. \& Piran T., 1996. ApJ, 1996, L41  
\bibitem{} Kouveliotou K., Meegan C. A., Fishman G. J. et al., 1993,
ApJ, 413, L101
\bibitem{} Gal--Yam A., et al., 2005, astro-ph/0509891
\bibitem{} Gehrels N. et al. 2004, ApJ, 611, 1005
\bibitem{} Ghirlanda G., Ghisellini G., Celotti A., 2004, A\&A, 422,
  L55 - GGC04
\bibitem{} Gladders et al., 2005, GCN, 3798
\bibitem{} Gorosabel J. \& Castro--Tirado A. J., 1997, ApJ, 483, L83
\bibitem{} Guetta D. \& Piran T., 2005, astro-ph/0511239
\bibitem{} Hamilton A. J. S., 1993, ApJ, 417, 19
\bibitem{} Hjorth J., et al., 2005, ApJ in press (astro-ph/0506123)
\bibitem{} Hurkett et al., 2005, GCN, 3381
\bibitem{} Hurley K. et al. 1999, ApJ, 515, 497
\bibitem{} Hurley K. et al., 2005, Nat, 434, 1098
\bibitem{} Lamb D. et al. 2004, NewAR, 48, 423
\bibitem[\protect\citename{Lilje }1988]{li} Lilje P.B., Efstathiou G.,
1988, MNRAS, 231, 635 
\bibitem{} Lazzati D., Ghirlanda G. \& Ghisellini G., 2005, MNRAS, 362, L8
\bibitem{} Ling E. N., Frenk C. S. \& Barrow J. D., 1986, MNRAS, 223, 21
\bibitem[\protect\citename{Madgwick1}2002]{Madwick1}
Madgwick D.S., et al. (the 2dFGRS Team), 2002, MNRAS, 333, 133
\bibitem[\protect\citename{Madgwick2}2002]{Madwick2}
Madgwick D.S., et al. (the 2dFGRS Team), 2003, MNRAS, 344, 847
\bibitem{} Magliocchetti M., Ghirlanda G., Celotti A., 2003, MNRAS,
  343, 255 (MGC03)
\bibitem{} Mazets E. P., astro-ph/0209219
\bibitem{} Nakar E., Gal--Yam A. \& Fox D. B., 2005, astro-ph/0511254
\bibitem{} Nakar E. et al., 2005, astro-ph/0502148
\bibitem{} Narayan R., Paczynski B., Piran T., 1992, ApJ, 395, L8
\bibitem[\protect\citename{Mo}1993]{Mo} Mo H.J., Peacock J.A., Xia,
X.Y., 1993, MNRAS, 260, 121
\bibitem{} Paciesas et al. 1999, ApJS, 122, 465
\bibitem[\protect\citename{Peebles }1974]{Pe}
Peebles P.J.E., 1974, ApJS, 28, 37
\bibitem[\protect\citename{Peebles1 }1980]{Pe}
Peebles P.J.E., 1980, ``The Large-Scale Structure of the Universe'', Princeton 
University Press. 
\bibitem{} Pedersen et al., 2005, ApJ in press, astro-ph/0510098
\bibitem{} Prochaska et al., 2005, GCN, 3679 
\bibitem{} Ruffert M., 1997, A\&A, 319, 122
\bibitem[\protect\citename{spe}2003]{Spe}
Spergel D.M., et al., 2003, ApJS, 148, 175
\bibitem{} Struble M. F. \& Rood H. J., 1997, ApJ, 490, 109
\bibitem{} Tanvir N. et al., 2005, Nature subm., astro-ph/0509167 
\bibitem[\protect\citename{tav}1998]{tav}
Tavani M., et al., 1998, ApJ, 497, L21
\bibitem{} Voss R., Tauris T. M., 2003, MNRAS, 342, 1169

\end{thebibliography}
\end{document}